\documentclass{amsart}
\usepackage{amsmath,amsthm,amssymb,IMjournal}

\usepackage[T1]{fontenc}
\usepackage{pifont}
\usepackage{amsfonts}
\usepackage{pstricks-add}
\usepackage{caption}
\usepackage{placeins}
\usepackage{pdfpages}
\usepackage{framed}
\usepackage{bigints}
\usepackage{bm}
\usepackage{epstopdf}
\usepackage{tikz}
\usepackage{enumitem}

\def\rr2dot{\mathop{\bf r}\limits}
\def\x2dot{\mathop{x}\limits}
\def\y2dot{\mathop{y}\limits}

\def\bfy2dot{\mathop{\bf y}\limits}
\def\z2dot{\mathop{z}\limits}

\def\csi2dot{\mathop{\xi}\limits}
\def\et2dot{\mathop{\eta}\limits}
\def\bet2dot{\mathop{\beta}\limits}
\def\t2dot{\mathop{\theta}\limits}
\def\s2dot{\mathop{\sigma}\limits}
\def\d2dot{\mathop{\delta}\limits}
\def\q2dot{\mathop{q}\limits}
\def\l2dot{\mathop{\lambda}\limits}
\def\ps2dot{\mathop{{\cal E}}\limits}
\def\tet2dot{\mathop{\theta}\limits}
\def\bfx2dot{\mathop{\bf x}\limits}
\def\bfy2dot{\mathop{\bf y}\limits}
\def\bfq2dot{\mathop{\bf q}\limits}
\def\bfr2dot{\mathop{\bf r}\limits}
\def\bbfq2dot{\mathop{\bar {\bf q}}\limits}
\def\w2{\mathop{W}\limits}
\def\xgrande2dot{\mathop{\bf X}\limits}
\def\p02dot{\mathop{P}\limits}
\def\a2dot{\mathop{A}\limits}

\begin{document}
\newtheorem{The}{Theorem}[section]
\newtheorem{rem}{Remark}[section]
\newtheorem{prop}{Proposition}[section]
\newtheorem{propr}{Property}[section]
\newtheorem{cor}{Corollary}[section]
\newtheorem{exe}{Example}[section]

\newcommand{\cal}{\mathcal}

\numberwithin{equation}{section}

\title{Transpositional rule for constrained systems}

\author{\|Federico |Talamucci|, Florence, Italy}

\rec {23 December, 2025}

\dedicatory{}\enddedicatory
 
\abstract
This paper investigates the dynamics of nonholonomic mechanical systems, focusing on fundamental variational assumptions and the role of the transpositional rule. We analyze how the $\check{\rm C}$etaev condition and the first variation of constraints define compatible virtual displacements for systems subject to kinematic constraints, including those nonlinear in generalized velocities. The study explores the necessary conditions for commutation relations to hold, clarifying their impact on the consistency of the derived equations of motion. By detailing the interplay between these variational identities and the Lagrangian derivatives of constraint functions, we elucidate the differences between equations of motion formulated via the d'Alembert--Lagrange principle and those obtained from extended time-integral variational principles. This work aims to provide a clearer theoretical framework for applying these core principles to nonholonomic dynamics.
\endabstract



\section{Introduction}\label{sec1}

\noindent
The dynamics of nonholonomic mechanical systems remains a foundational challenge in analytical mechanics. Unlike holonomic systems, where integrable constraints allow for a straightforward reduction of degrees of freedom, nonholonomic systems involve non-integrable kinematic constraints dependent on generalized velocities. This non-integrability requires a re-evaluation of fundamental principles, particularly concerning virtual variations and their compatibility with motion restrictions. Consequently, the formulation of equations of motion for these systems has historically sparked academic debate, resulting in diverse and sometimes conflicting theoretical approaches.

\noindent
This study provides a cohesive understanding of these systems by focusing on the transpositional rule. This identity serves as a bridge between various conventions used to handle variations in nonholonomic dynamics. We analyze how this rule, alongside $\check{\rm C}$etaev's condition for ideal virtual displacements and the first variation of constraints, influences the formulation of the equations of motion. Our investigation explores the interplay between these principles and the implications for the resulting dynamic models.

\noindent
A central theme is the analysis of the commutation relations between variational and time-derivative operators. We clarify how different assumptions regarding these relations are not merely formal, but fundamentally affect the consistency and physical validity of the derived equations. By comparing the d'Alembert--Lagrange principle with generalized time-integral variational principles—such as the Hamilton--Suslov principle and vakonomic mechanics—this work highlights critical distinctions and seeks to offer a more unified perspective on nonholonomic dynamics.

\subsection{The physical framework}

\noindent
Consider a mechanical system described by ge\-ne\-ra\-li\-zed coordinates ${\bf q}=(q_1, \dots, q_n)$ and velocities ${\dot {\bf q}}=({\dot q}_1, \dots, {\dot q}_n)$, subject to $\kappa < n$ nonholonomic constraints
\begin{equation}\label{vincg}g_\nu ({\bf q}, {\dot {\bf q}}, t)=0, \qquad \nu=1, \dots, \kappa.\end{equation}
The constraints are assumed to be regular, satisfying the non-singularity condition:
\begin{equation}\label{gind}\text{rank} \left(\dfrac{\partial (g_1, \dots, g_\kappa)}{\partial ({\dot q}_1, \dots, {\dot q}_n)}\right)=\kappa.
\end{equation}
The functions $g_\nu$ can be linear or nonlinear in the velocities ${\dot q}_i$. In the linear case, \eqref{vincg} takes the form
\begin{equation}
\label{glin}g_\nu ({\bf q}, {\dot {\bf q}},t)=\sum\limits_{j=1}^n a_{\nu,j}({\bf q,t}){\dot q}_j+b_\nu({\bf q},t).
\end{equation}
A constraint is integrable (holonomic) if there exists a function $f_\nu ({\bf q},t)$ such that:\begin{equation}\label{gint}g_\nu ({\bf q}, {\dot {\bf q}},t)=\dfrac{d}{dt}f_\nu ({\bf q},t)=\sum\limits_{i=1}^n \dfrac{\partial f_\nu}{\partial q_i}{\dot q}_i+\dfrac{\partial f_+\nu}{\partial t}.\end{equation}

\noindent
Condition \eqref{gind} allows us to express \eqref{vincg} in explicit form. By re-indexing the variables if necessary to ensure that the Jacobian of the constraints with respect to $({\dot q}_{m+1}, \dots, {\dot q}_n)$ is non-zero, we can write
\begin{equation}
\label{constrexpl}
{\dot q}_{m+\nu} = \alpha_\nu(q_1, \dots, q_n, {\dot q}_1, \dots, {\dot q}_m, t), \qquad \nu=1, \dots, \kappa
\end{equation}
where $m = n - \kappa$. In this formulation, $({\dot q}_1, \dots, {\dot q}_m)$ serve as independent kinetic variables, and the corresponding virtual displacements $(\delta q_1, \dots, \delta q_m)$ are likewise independent.

\subsection{Variations and transpositional rule}

\subsubsection{First variation and ${\check{\rm C}}$etaev condition}

\noindent
We distinguish two types of variations for a differentiable function $F({\bf q}, {\dot {\bf q}},t)$: the $\check{\rm C}$etaev variation $\delta^{(c)}F$ and the total variation $\delta^{(v)}F$, defined respectively as
\begin{equation}
\label{dcvdef}
\delta^{(c)}F:=\sum\limits_{i=1}^n \dfrac{\partial F}{\partial {\dot q}_i}\delta q_i, \qquad
\delta^{(v)}F:=\sum\limits_{i=1}^n \dfrac{\partial F}{\partial q_i}\delta q_i+ 
\sum\limits_{i=1}^n \dfrac{\partial F}{\partial {\dot q}_i}\delta {\dot q}_i.
\end{equation}
The condition $\delta^{(v)}g_\nu = 0$ is generally required to derive equations of motion from time-integral variational principles. Conversely, the assumption $\delta^{(c)} g_\nu = 0$, known as the $\check{\rm C}$etaev condition \cite{cetaev}, defines the constraints as ideal or perfect.

\noindent
Although primarily a postulate, the $\check{\rm C}$etaev condition effectively generalizes holonomic theory. For integrable constraints $g_\nu = \frac{d}{dt}f_\nu$, it reduces to $\sum \frac{\partial f_\nu}{\partial q_i}\delta q_i = 0$, matching the standard definition of ideal geometric constraints. For linear kinematic constraints \eqref{glin}, it correctly reproduces the classical virtual displacement condition $\sum a_{\nu,j}\delta q_j = 0$ \cite{neimark}.

\subsubsection{The transpositional rule}

\noindent
The variations $\delta^{(c)}$ and $\delta^{(v)}$ are linked by the transpositional rule
\begin{equation}
\label{transprule}
	\delta^{(v)} F -\dfrac{d}{dt}\left(\delta^{(c)}F\right)=\sum\limits_{i=1}^n \dfrac{\partial F}{\partial {\dot q}_i}\left(\delta {\dot q}_i - \dfrac{d}{dt}\delta q_i\right)-\sum\limits_{i=1}^n {\cal D}_i F \delta q_i
\end{equation}
where ${\cal D}_i F = \frac{d}{dt}\frac{\partial F}{\partial {\dot q}_i} - \frac{\partial F}{\partial q_i}$ denotes the $i$-th Lagrangian derivative. The term $(\delta {\dot q}_i - \frac{d}{dt}\delta q_i)$ accounts for the potential non-commutation of the variation and time-derivative operators. The assumption $\delta {\dot q}_i = \frac{d}{dt}(\delta q_i)$ defines the standard commutation relations.

\noindent
Applying \eqref{transprule} to the constraint functions $g_\nu$ allows for a rigorous comparison between the conditions $\delta^{(c)}g_\nu=0$ and $\delta^{(v)}g_\nu=0$, highlighting their dependence on the commutation relations and the Lagrangian derivatives of the constraints.

\subsubsection{Transpositional rule for explicit constraints}

\noindent
For constraints in the explicit form \eqref{constrexpl}, the variations and Lagrangian derivatives 
become
\begin{eqnarray}
	\nonumber
& &\delta^{(c)}g_\nu = \delta q_{m+\nu} - \sum\limits_{r=1}^m \dfrac{\partial \alpha_\nu}{\partial {\dot q}r}\delta q_r, \quad
\delta^{(v)}g_\nu = \delta {\dot q}_{m+\nu} - \sum\limits_{i=1}^n \dfrac{\partial \alpha_\nu}{\partial q_i}\delta q_i - \sum\limits_{r=1}^m \dfrac{\partial \alpha_\nu}{\partial {\dot q}_r}\delta {\dot q}_r 
\\
& &\label{dexpl}
\\
\nonumber
& &{\cal D}_i g_\nu = 
\left\{\
\begin{array}{ll} 
	-{\cal D}_i\alpha_\nu  &\text{for } i=1, \dots, m \\ 
	\dfrac{\partial \alpha_\nu}{\partial q_i}  &\text{for } i=m+1, \dots, n 
\end{array}
\right.
\end{eqnarray}
Substituting these into \eqref{transprule} yields the explicit transpositional rule
\begin{eqnarray}
\nonumber\delta^{(v)} g_\nu -\dfrac{d}{dt}\left(\delta^{(c)}g_\nu\right) 
&=&\left[ \delta {\dot q}_{m+\nu}-\dfrac{d}{dt}( \delta q_{m+\nu}) \right]-\sum\limits_{r=1}^m \dfrac{\partial \alpha_\nu}{\partial {\dot q}r}\left(\delta {\dot q}_r - \dfrac{d}{dt}\delta q_r\right)
\\ 
\label{transpruleexpl}
&+& \sum\limits_{r=1}^m {\cal D}_r \alpha_\nu \delta q_r - \sum\limits_{\mu=1}^\kappa \dfrac{\partial \alpha_\nu}{\partial q_{m+\mu}}\delta q_{m+\mu}.
\end{eqnarray}
Under this framework, the $\check{\rm C}$etaev condition $\delta^{(c)}g_\nu = 0$ and the variation $\delta^{(v)}g_\nu = 0$ define the non-independent variations $\delta q_{m+\nu}$ and $\delta \dot{q}_{m+\nu}$ in terms of the independent variables $(q_r, \dot{q}_r)$.


\subsection{Equations of motion for nonholonomic systems: an overview}

\noindent
To cla\-ri\-fy the roles of the variations \eqref{dcvdef} and of the transpositional rule \eqref{transprule}, we briefly outline two primary approaches to nonholonomic dynamics: the d'Alembert--Lagrange principle and generalized time-integral variational principles.

\subsubsection{The d'Alembert--Lagrange principle}

\noindent
This principle states that
\begin{equation}
\label{dal}\sum\limits_{i=1}^n \left( \dfrac{d}{dt}\dfrac{\partial {\cal L}}{\partial {\dot q}_i}-\dfrac{\partial {\cal L}}{\partial q_i}\right) \delta q_i=0
\end{equation}for any virtual displacement $\delta {\bf q}$ compatible with the constraints \eqref{vincg}. Using the $\check{\rm C}$etaev condition (first condition in \eqref{dcvdef}), the variations $\delta {\bf q}$ are restricted to the subspace orthogonal to $\text{span}(\nabla_{\dot{\bf q}} g_\nu)$. This leads to the equations of motion with Lagrange multipliers $\mu_\nu$
\begin{equation}
\label{eqmotodc}
\dfrac{d}{dt}\dfrac{\partial {\cal L}}{\partial {\dot q}_i}-
\dfrac{\partial {\cal L}}{\partial q_i}=\sum\limits_{\nu=1}^\kappa \mu_\nu 
\dfrac{\partial g_\nu}{\partial {\dot q}_i}, \qquad i=1, \dots, n.
\end{equation}
Notably, these equations can be derived without explicit assumptions regarding the commutation relations $\delta {\dot q}_i=\frac{d}{dt}(\delta q_i)$, $i=1, \dots, n$.

\subsubsection{The equations of motion via a time-integral variational principle}

\noindent
Extending Hamilton’s principle to nonholonomic systems remains an open and debated issue \cite{cronstrom, flanneryenigma, krup4, lemos, mei, rumy2000}. To address this, two primary modifications are typically introduced:
\begin{itemize}
	\item[1.]
Constrained Action: the standard stationarity condition is replaced by a constrained functional using Lagrange multipliers $\lambda_\nu$:
\begin{equation} 
\label{hsp}\delta \int_{t_0}^{t_1} \left({\cal L} + \sum_{\nu=1}^\kappa \lambda_\nu g_\nu \right) dt = 0.\end{equation}
\item[2.]
Non-commutation: relaxing the standard commutation relations leads to the Hamilton–Suslov principle \cite{neimark}, expressed as
\begin{equation} 
\label{sa}\int_{t_0}^{t_1} \left( \delta L - \sum_{i=1}^n \dfrac{\partial L}{\partial {\dot q}_i} \left(\delta {\dot q}_i - \dfrac{d}{dt} \delta q_i \right) \right) dt = 0,
\end{equation}
where the term $\delta {\dot q}_i - \frac{d}{dt}\delta q_i$ accounts for the variation of the velocity.
\end{itemize}

\noindent
By combining these aspects and defining the constrained Lagrangian ${\cal L}_R = {\cal L} + \sum \lambda_\nu g_\nu$, the principle takes the form
\begin{equation} 
\label{hs}\int_{t_0}^{t_1} \left( \delta {\cal L}_R - \sum\limits_{i=1}^n \dfrac{\partial {\cal L}_R}{\partial {\dot q}_i} \left(\delta {\dot q}_i - \dfrac{d}{dt} \delta q_i \right) \right) dt = 0.
\end{equation}
If the non-commutation (transpositional rules) is modeled via coefficients $W_{i,j}$ such that 
\begin{equation}
\label{tra}
\delta {\dot q}_i - \frac{d}{dt}\delta q_i = \sum W_{i,j} \delta q_j, \quad i=1, \dots, n
\end{equation}
one derives the modified vakonomic equations \cite{llibre, pastore}:
\begin{equation} 
\label{eqllibre}{\cal D}_i{\cal L}_R - \sum\limits_{j=1}^n W{j,i}\dfrac{\partial {\cal L}_R}{\partial {\dot q}_j} = 0, \quad i=1, \dots, n.
\end{equation}
Together with constraints $g_\nu=0$, these form a system of $n+\kappa$ equations. 

\noindent
Special cases and assumptions:
\begin{itemize}
	\item{}
Standard Vakonomic Mechanics: if $W_{i,j}=0$, eq. \eqref{eqllibre} reduces to the classical vakonomic form \cite{arnoldkozlov, ray}
\begin{equation} 
\label{eqray}\dfrac{d}{dt}\dfrac{\partial {\cal L}}{\partial {\dot q}i} - \dfrac{\partial {\cal L}}{\partial q_i} = \sum\limits_{\nu=1}^\kappa \left( \lambda_\nu \left( \dfrac{\partial g_\nu}{\partial q_i} - \dfrac{d}{dt}\dfrac{\partial g_\nu}{\partial {\dot q}i} \right) - {\dot \lambda}\nu \dfrac{\partial g_\nu}{\partial {\dot q}_i} \right).
\end{equation}
This formulation differs from the d'Alembert-Lagrange dynamics \eqref{eqmotodc}, though both recover standard Lagrange equations in the absence of constraints.
\item{}
Operative Assumption: following \cite{llibre}, if one assumes 
\begin{equation}
\label{llibre2}
{\cal D}_i g_\nu - \sum\limits_{j=1}^n W_{j,i}\frac{\partial g_\nu}{\partial {\dot q}_j} = 0, \quad i=1, \dots, n \quad \nu=1, \dots, \kappa
\end{equation}
the equations of motion simplify to
\begin{equation} 
\label{eqllibre2}
{\cal D}_i{\cal L} - \sum\limits_{j=1}^n W_{j,i}\dfrac{\partial {\cal L}}{\partial {\dot q}_j} = -\sum\limits_{j=1}^n \sum_{\nu=1}^\kappa W_{j,i} {\dot \lambda}\nu \dfrac{\partial g\nu}{\partial {\dot q}_j}, \quad i=1, \dots, n.
\end{equation}
\end{itemize}

\noindent
The determination of $W_{i,j}$ remains the central theoretical challenge in this framework \cite{pastore}.

\section{The mathematical aspect}

\noindent
To clarify the subsequent analysis, we summarize the fundamental conditions:
\begin{equation}
\label{abc0}
\begin{array}{ll}
(A) & \delta^{(c)} g_\nu = \sum\limits_{i=1}^n \dfrac{\partial g_\nu}{\partial {\dot q}_i}\delta q_i=0 \\
\\
(B) & \delta^{(v)} g_\nu= \sum\limits_{i=1}^n \dfrac{\partial g_\nu}{\partial q_i}\delta q_i+ \sum_{i=1}^n \dfrac{\partial g_\nu}{\partial {\dot q}_i}\delta {\dot q}_i=0 \\
\\
(C_0) & \delta {\dot q}_i - \dfrac{d}{dt}(\delta q_i)=0, \quad i=1, \dots, n \\
\\
(C) & \delta {\dot q}_i - \dfrac{d}{dt}(\delta q_i) = \sum\limits_{j=1}^n W_{i,j}\delta q_j, \quad i=1, \dots, n \\
\\ 
(C_1) & \delta {\dot q}_r - \dfrac{d}{dt}(\delta q_r)=0, \quad r=1, \dots, m
\end{array}
\end{equation}
where $(C_1)$ assumes commutation holds only for independent variables. 

\noindent
The specific dynamical approach is defined by the choice of these conditions. For instance, while $(C_1)$ assumes that commutation holds only for independent variables, the $\check{\rm C}$etaev formulation \eqref{eqmotodc} relies solely on $(A)$. In contrast, the combination of $(A)$ and $(C_0)$ corresponds to the Hölder principle \cite{rumycet}. Similarly, the coupling of $(B)$ and $(C_0)$ underpins the vakonomic equations of motion \eqref{eqray} \cite{arnoldkozlov}, whereas the broader assumption $(C)$, in conjunction with $(B)$, leads to recent generalizations of vakonomic mechanics \cite{llibre}, \cite{pastore}.

\subsection{Necessary conditions for commutation}

\noindent
The transpositional rule \eqref{transprule} reveals the necessary conditions for commutation under various assumptions:
\begin{prop}
If the commutation relations $(C_0)$ hold, then:
\begin{itemize}
\item Under $(A)$: $\delta^{(v)}g_\nu = -\sum\limits_{i=1}^n {\cal D}_ig_\nu \delta q_i$,
\vspace{1truemm}
\item Under $(B)$: $\dfrac{d}{dt}(\delta^{(c)}g_\nu) = \sum\limits_{i=1}^n {\cal D}_i g_\nu \delta q_i$,
\vspace{1truemm}
\item Under $(AB)$: $\sum\limits_{i=1}^n {\cal D}_i g_\nu \delta q_i = 0$.
\end{itemize}
\end{prop}

\noindent
Notably, if 
\begin{equation}
	\label{derlagrsomma0}
	\sum\limits_{i=1}^n {\cal D}_ig_\nu \delta q_i=0, \quad \nu=1, \dots, \kappa.
\end{equation}
is satisfied, then \eqref{transprule} implies $\delta^{(v)} g_\nu = \frac{d}{dt}(\delta^{(c)}g_\nu)$. In this case, $(A)$ and $(B)$ become equivalent.

\noindent
The converse of the Proposition does not hold: while any of the three conditions implies that
$\sum_{i=1}^n \frac{\partial g_\nu}{\partial \dot{q}_i} \left( \delta \dot{q}_i - \frac{d}{dt} \delta q_i \right) = 0$
for all $\nu = 1, \dots, \kappa$, this does not in turn imply $(C_0)$, despite the fact that $\kappa <n$.

\subsection{The double assumption $(A)$ and $(B)$}

\noindent
Assumption $(A)$ serves a physical role, characterizing ideal constraints through virtual displacements. In contrast, assumption $(B)$ pertains primarily to the implementation of variational principles and the formal techniques of the calculus of variations.
The combined assumption $(AB)$, widely discussed in the literature, is common in both classical tratises \cite{lurie, pars} and contemporary models. However, there is widespread skepticism regarding the coupling $(AB)$ when applied within the framework of the commutation relations $(C_0)$; it is often argued that, in this case, only integrable constraints can be considered. According to recent approaches \cite{llibre, pastore}, replacing $(C_0)$ with a more general assumption $(C)$ (see (\ref{abc0})) the transpositional rule (\ref{transprule}) allows $(AB)$ and $(C)$ to coexist even in the context of nonholonomic systems. Specifically, we address the following question: assuming $(AB)$, what conclusions can be drawn regarding properties $(C_0)$ and (\ref{derlagrsomma0})? An illustration of the case under consideration is provided below:
\begin{prop}
Assume $(AB)$. If the commutation relations $(C_0)$ are verified, then $\sum\limits_{i=1}^n {\cal D}_ig_\nu \delta q_i = 0$.
\end{prop}

\noindent
Indeed, if $(A)$ and $(B)$ are both in force then the transpositional rule reduces to 
\begin{equation}
	\label{transpruleab}	
	\sum\limits_{i=1}^n 
	\dfrac{\partial g_\nu}{\partial {\dot q}_i}\left(\delta {\dot q}_i - \dfrac{d}{dt}\delta q_i\right)=\sum\limits_{i=1}^n {\cal D}_i g_\nu \delta q_i
\end{equation}
and the sum on the left side is zero if $(C_0)$ is verified, whence (\ref{derlagrsomma0}) must hold.

\noindent
Under $(AB)$, the identity \eqref{derlagrsomma0} is a necessary condition for $(C_0)$. Similarly to a previous remark, the same identity does not imply the commutation $(C_0)$, but only 
$\sum\limits_{i=1}^n \frac{\partial g_\nu}{\partial {\dot q}_i}\left(\delta {\dot q}_i 
- \frac{d}{dt}\delta q_i\right)=0$.

\noindent
Under the sole assumption of $(A)$, we can prove a useful characterization of condition \eqref{derlagrsomma0}:

\begin{prop}
Assuming $\delta^{(c)}g_\nu=0$ for $\nu=1, \dots, \kappa$, the equality (\ref{derlagrsomma0}) holds if and only if
\begin{equation}
\label{zero2}{\cal D}_i g_\nu =\sum_{\mu=1}^\kappa\varrho_\mu^{(\nu)}\dfrac{\partial g_\mu}{\partial {\dot q}_i} \qquad (i=1, \dots, n)
\end{equation}
for some coefficients $\varrho\mu^{(\nu)}({\bf q}, {\dot {\bf q}},t)$.
\end{prop}

\noindent{\bf Proof}.If (\ref{derlagrsomma0}) holds for every displacement satisfying $(A)$, then (\ref{zero2}) follows immediately from the full--rank condition (\ref{gind}).Conversely, multiplying (\ref{zero2}) by $\delta q_i$ and summing over $i$ yields
$$
\sum_{i=1}^n {\cal D}_i g_\nu \delta q_i = \sum_{\mu=1}^\kappa\varrho_\mu^{(\nu)} \left( \sum_{i=1}^n \dfrac{\partial g_\mu}{\partial {\dot q}_i}\delta q_i \right) = 0
$$
which vanishes by virtue of condition $(A)$.$\quad \square$

\noindent
The equivalent condition (\ref{zero2}) suggests that only linear constraints can strictly satisfy $(C_0)$, as nonlinear constraints would introduce second-order derivatives $\ddot{q}_i$ in the Lagrangian derivative ${\cal D}_i g_\nu$.

\subsubsection{Geometric structure of displacements}

\noindent
In vector-matrix notation, let $A = [\partial g_\nu / \partial q_j]$ and $B = [\partial g_\nu / \partial \dot{q}_j]$, $\nu=1, \dots, \kappa$, $j=1, \dots, n$
and let $\delta {\bf q}, \delta {\dot {\bf q}} \in \mathbb{R}^n$ be the variation column vectors. 
The conditions $(AB)$ correspond to the system
\begin{equation}
	\label{abvett}
	\left\{
	\begin{array}{ll}
		B\delta {\bf q}={\bf 0}& \qquad \qquad (A)\\
		\\
		A\delta {\bf q}+B\delta {\dot {\bf q}}={\bf 0}& \qquad \qquad (B)\\
	\end{array}
	\right.
\end{equation}
The main results are summarized in Propositions 2.4 and 2.5, which characterize the linear spaces containing the displacements $\delta {\bf q}$ and variations ${\delta {\dot \bf q}}$ under the condition $(AB)$.

\begin{prop}
Under $(AB)$, the set of consistent virtual displacements $\delta {\bf q}$ is the $(n-\kappa)$-dimensional subspace ${\Bbb V}^{(n-\kappa)} = \ker B$. 
For a fixed $\delta {\bf q}$, the set of consistent variations $\delta \dot{\bf q}$ is an affine space ${\Bbb A}^{(n-\kappa)}$ defined by
\begin{equation}
\label{deltadotq}\delta {\dot {\bf q}}=\sum_{s=1}^m \psi_s {\bf w}_s - B^+A \delta {\bf q}\end{equation}where $\{{\bf w}_s\}$ is a basis for $\ker B$ and $B^+$ is a right inverse of $B$.
\end{prop}

\noindent{\bf Proof}.
Given that $\text{rank}\,B=\kappa$, the solution space of $(A)$ is the kernel of $B$, which is the orthogonal complement of the space spanned by the rows of $B$:
\begin{equation}
\label{w}
{\Bbb V}^{(n-\kappa)} = \ker B = \left\langle \dfrac{\partial g_1}{\partial {\dot {\bf q}}}, \dots, \dfrac{\partial g_\kappa}{\partial {\dot {\bf q}}}\right\rangle^\perp.
\end{equation}
Thus, $\delta {\bf q} = \sum_{r=1}^m \sigma_r {\bf w}_r$ proves the first part.For a fixed $\delta {\bf q}$, equation $(B)$ is a non-homogeneous linear system for $\delta {\dot {\bf q}}$. Its general solution is
\begin{equation}
\label{deltadotq_sol}\delta {\dot {\bf q}}=\sum_{s=1}^m \psi_s {\bf w}s - B^+A \delta {\bf q}
\end{equation}
where $B^+ \in \mathbb{R}^{n \times \kappa}$ is a right inverse of $B$, that is $BB^+={\Bbb I}_\kappa$ (we recall that $rank B=\kappa$ and existence of $B^+$ are equivalent conditions and that $B^+{\bf b}$ is a particular solution of the system $B{\bf x}={\bf b}$, see \cite{golub}). The first term spans $\ker B$ (the space ${\Bbb V}^{(n-\kappa)}$), while the second is a particular solution. This shows that the set of compatible $\delta {\dot {\bf q}}$ is the affine space ${\Bbb A}^{(n-\kappa)}$ resulting from the translation of ${\Bbb V}^{(n-\kappa)}$. $\quad \square$

\noindent
Using the matrices $A$ and $B$ we define $D = \dot{B} - A \in \mathbb{R}^{\kappa \times n}$. 
The relation (\ref{transpruleab}) can be written in vector-matrix form as:
\begin{equation}
\label{transpruleabmatr}
B\left(\delta {\dot {\bf q}}-\dfrac{d}{dt}(\delta {\bf q})\right)=D\delta {\bf q}.
\end{equation}

\begin{prop}Assume $(AB)$. Then:
\begin{enumerate}[label=(\roman*)]\item $(C_0)$ implies $D \delta {\bf q}={\bf 0}$.\item If $D {\delta {\bf q}}={\bf 0}$, then $\delta {\dot {\bf q}}-\frac{d}{dt}(\delta {\bf q}) \in \ker B$.\item If $D {\delta {\bf q}}={\bf 0}$, then $\frac{d}{dt} (\delta {\bf q})\in {\Bbb A}^{(n-\kappa)}$, where ${\Bbb A}^{(n-\kappa)}$ is the affine space defined in (\ref{deltadotq}).
\end{enumerate}
\end{prop}

\noindent{\bf Proof}. Property $(i)$ follows directly from (\ref{transpruleabmatr}) by substituting the condition $(C_0): \delta \dot{\bf q} = \frac{d}{dt}(\delta {\bf q})$. To prove $(ii)$, the assumption $D\delta {\bf q}={\bf 0}$ reduces (\ref{transpruleabmatr}) to $B(\delta {\dot {\bf q}}-\frac{d}{dt}\delta {\bf q})={\bf 0}$, which by definition implies that the vector difference lies in $\ker B = \text{span}\langle {\bf w}_1,\dots,{\bf w}_m\rangle$. For $(iii)$, we differentiate the consistency condition $(A)$, $B \delta {\bf q} = {\bf 0}$, with respect to time:
\begin{equation}
\label{diffA}
\dot{B}\delta {\bf q} + B \dfrac{d}{dt}(\delta {\bf q}) = {\bf 0} \implies B \dfrac{d}{dt}(\delta {\bf q}) = -\dot{B}\delta {\bf q}.
\end{equation}
Under the assumption $D\delta {\bf q} = {\bf 0}$ (i.e., $A\delta {\bf q} = \dot{B}\delta {\bf q}$), equation (\ref{diffA}) becomes $B \frac{d}{dt}(\delta {\bf q}) = -A\delta {\bf q}$. This is exactly the non-homogeneous system (\ref{abvett}) defining the affine space ${\Bbb A}^{(n-\kappa)}$ for $\delta \dot{\bf q}$. Thus, $\frac{d}{dt}(\delta {\bf q}) \in {\Bbb A}^{(n-\kappa)}$. $\quad\square$

\subsubsection{Remarks on Proposition 2.5} 

\noindent
The condition $D\delta {\bf q}={\bf 0}$ is necessary, but not sufficient, for the commutation rule $(C_0)$ to hold. Physically, $D\delta {\bf q} = {\bf 0}$ restricts the difference $\delta {\dot {\bf q}}-\frac{d}{dt}(\delta {\bf q})$ to $\ker B$. Geometrically, this implies that if $D\delta {\bf q} \neq {\bf 0}$, the affine spaces for $\delta \dot{\bf q}$ and $\frac{d}{dt}(\delta {\bf q})$ are disjoint, as the underlying linear systems share the same matrix $B$ but have distinct non-homogeneous terms ($-A\delta {\bf q} \neq -\dot{B}\delta {\bf q}$). Regarding the broader assumption $(C)$ in (\ref{tra}), often employed to extend vakonomic techniques to nonholonomic systems, we compare our approach with the $\kappa \times n$ conditions (\ref{llibre2}) proposed in \cite{llibre}. While (\ref{llibre2}) implies the transpositional rule (\ref{transpruleab}) via the identity
\begin{equation}
\label{llibredq_rev}
\sum_{i=1}^n {\cal D}_i g_\nu\delta q_i - \sum_{j=1}^n \dfrac{\partial g_\nu}{\partial {\dot q}_j}\left( \delta {\dot q}_j - \dfrac{d}{dt}(\delta q_j)\right) = 0, \quad \nu=1, \dots, \kappa,
\end{equation}
the converse is generally false. Specifically, (\ref{transpruleab}) imposes only $n$ constraints on the coefficients $W{i,j}$, whereas (\ref{llibre2}) requires $\kappa \times n$ conditions. Consequently, the relations in \cite{llibre} constitute additional hypotheses which appear as restrictive requirements rather than necessary consequences of the dynamics.

\subsubsection{Explicit constraints and assumption $(C_1)$ (partial commutation)}

For systems with explicit constraints \eqref{constrexpl}, assumptions $(A)$ and $(B)$ yield the transpositional relation (see (\ref{dexpl}) and (\ref{transpruleexpl}))
\begin{equation}
\label{transpruleexplab}
\delta {\dot q}_{m+\nu}-\dfrac{d}{dt} \delta q_{m+\nu}=\sum\limits_{r=1}^m \dfrac{\partial \alpha_\nu}{\partial {\dot q}_r}\left(\delta {\dot q}r - \dfrac{d}{dt}\delta q_r\right)
-\sum\limits_{r=1}^m \left( {\cal D}_r \alpha_\nu-\sum\limits_{\mu=1}^\kappa \dfrac{\partial \alpha_\nu}{\partial q_{m+\mu}}\dfrac{\partial \alpha_\mu}{\partial {\dot q}_r}\right)\delta q_r,
\end{equation}
where $\nu=1, \dots, \kappa$ and $\alpha_\nu = \alpha_\nu(q, \dot{q}_1, \dots, \dot{q}_m, t)$. If we further assume $(C_1)$, that is the commutation relations hold only for the $m$ independent variables, equation \eqref{transpruleexplab} simplifies to
\begin{equation}
\label{treabc1}
\delta {\dot q}_{m+\nu}-\dfrac{d}{dt} \delta q_{m+\nu}= -\sum_{r=1}^m \left( {\cal D}_r \alpha_\nu-\sum_{\mu=1}^\kappa \dfrac{\partial \alpha_\nu}{\partial q_{m+\mu}}\dfrac{\partial \alpha_\mu}{\partial {\dot q}_r}\right)\delta q_r.
\end{equation}
This framework, rooted in the Hamilton–Suslov principle \cite{neimark} and widely adopted \cite{papa, rumy2000}, leads to the following result:

\begin{prop}
Under assumptions $(A)$, $(B)$ and $(C_1)$, the commutation relations $\delta \dot{q}_{m+\nu} = \frac{d}{dt} \delta q_{m+\nu}$ hold for the dependent variables if and only if
\begin{equation}
\label{derlagrsomma0}
\sum_{i=1}^n {\cal D}_i g_\nu \delta q_i = 0, \quad \nu = 1, \dots, \kappa.
\end{equation}
\end{prop}

\noindent{\bf Proof}. The result follows directly from \eqref{treabc1} by noting that its right-hand side is equivalent to $-\sum_{i=1}^n {\cal D}_i g_\nu \delta q_i$ (see \eqref{transpruleexpl}). $\square$

\subsubsection{Historical Context and $(C_1)$ as a Hypothesis}

\noindent
The choice of commutation rules defines different approaches to nonholonomic mechanics. The global commutation $(C_0)$ for all variables, supported by Volterra, Hamel, and Pars \cite{neimark, pars}, implies $d/dt$ and $\delta$ commute throughout. Conversely, the partial commutation $(C_1)$, restricted to independent variables, was introduced by Suslov, Levi-Civita and Amaldi. While $(C_1)$ simplifies the link between global commutation and the condition ${\cal D}_i g_\nu \delta q_i = 0$, it remains a hypothesis. In its absence, the resulting variational principles and the correspondence between different formulations of the equations of motion become significantly weaker.


\section{Categories of nonholonomic constraints}

We analyze special classes of constraints, distinguishing between linear and nonlinear formulations with respect to generalized velocities.

\subsection{Linear Kinematic Constraints}
Consider constraints of the form $g_\nu({\bf q}, \dot{\bf q}, t) = \sum a_{\nu,i}\dot{q}_i + b_\nu = 0$. The transpositional rule \eqref{transprule} is expressed as
\begin{equation}
\label{transprellin}
\delta^{(v)}g_\nu - \dfrac{d}{dt}\delta^{(c)}g_\nu = \sum_{j=1}^n \dfrac{\partial g_\nu}{\partial {\dot q}_j} \left(\delta {\dot q}_j - \dfrac{d}{dt}\delta q_j\right) - \sum{j=1}^n D_j g_\nu \delta q_j.
\end{equation}
Under the $\check{\rm C}$etaev condition (Assumption $A$), $\delta^{(c)}g_\nu = 0$, this formulation overlaps with the standard treatment of virtual displacements \cite{neimark}

\subsubsection{Explicit Form and Commutation}
When constraints are given explicitly as $\dot{q}_{m+\nu} = \sum_{r=1}^m \xi_{\nu,r}\dot{q}_r + \eta_\nu$, the transpositional rule becomes
\begin{eqnarray}
\nonumber
\delta^{(v)}g_\nu - \dfrac{d}{dt}\delta^{(c)}g_\nu &=& \left[\delta {\dot q}_{m+\nu}-\dfrac{d}{dt} \delta q_{m+\nu}\right] - \sum\limits_{r=1}^m \xi_{\nu,r}\left(\delta {\dot q}_r - \dfrac{d}{dt}\delta q_r\right)\\
\label{trlinexpl}
&-& \sum\limits_{r,s=1}^m \beta_{sr}^{m+\nu}{\dot q}_s \delta q_r - \sum\limits_{r=1}^m \gamma_r^\nu \delta q_r.
\end{eqnarray}
The coefficients $\beta_{sr}^{m+\nu}$ and $\gamma_r^\nu$ account for the non-integrability and time-dependence of the constraints. If Assumption $(C_1)$ (commutation for independent variables) holds, the first summation vanishes. For autonomous constraints ($\eta_\nu=0, \partial_t \xi = 0$), $\gamma_r^\nu = 0$.

\subsubsection{Linear Homogeneous Constraints}
For constraints $g_\nu = \sum a_{\nu,j}({\bf q})\dot{q}_j = 0$, the condition $\sum\limits_{j=1}^n D_j g_\nu \delta q_j = 0$ is central. In vector notation, this requires
\begin{equation}
\label{conj}
\dot{\bf q}^T \mathbb{A} \delta {\bf q} = 0, \quad \text{where } \mathbb{A} = \left(\dfrac{\partial {\bf a}\nu}{\partial {\bf q}}\right)^T - \dfrac{\partial {\bf a}\nu}{\partial {\bf q}}.
\end{equation}
Since $\mathbb{A}$ is skew-symmetric, the product is zero if $\dot{\bf q}$ and $\delta{\bf q}$ are parallel.

\noindent
Case $n=2$: Since $\dot{\bf q}$ and $\delta{\bf q}$ both belong to the 1D space orthogonal to ${\bf a}_\nu$, they are necessarily parallel; thus \eqref{conj} is always satisfied.

\noindent
Case $n>2$: The orthogonal space has dimension $\geq 2$, so $\dot{\bf q}$ and $\delta{\bf q}$ are not necessarily parallel, and \eqref{conj} may not vanish.

\subsubsection{Exact constraints}

\noindent
An exact (or integrable) constraint $g_\nu$ satisfies (\ref{gint}) for some function $f_\nu({\bf q},t)$. Such a constraint is linear, with coefficients $a_{\nu,j}=\frac{\partial f_\nu}{\partial q_j}$ and $b_\nu = \frac{\partial f_\nu}{\partial t}$.
The relations $\frac{\partial {\dot f}_\nu}{\partial {\dot q}_i}=\frac{\partial f_\nu}{\partial q_i}$, $\frac{d}{dt}\frac{\partial f_\nu}{\partial q_i}=\frac{\partial {\dot f}_\nu}{\partial q_i}$
imply $D_i{\dot f}_\nu = \frac{d}{dt}\frac{\partial {\dot f}_\nu}{\partial {\dot q}_i} - \frac{\partial {\dot f}_\nu}{\partial q_i} = 0$.
Consequently, the transpositional rule becomes
\begin{equation}
\label{transprulee}\delta^{(v)} {\dot f}_\nu -\frac{d}{dt}\left( \delta^{(c)}{\dot f}_\nu \right) = \sum\limits_{j=1}^n \frac{\partial f_\nu}{\partial q_j}\left(\delta {\dot q}_j - \frac{d}{dt}\delta q_j\right).
\end{equation}
Since the ${\check{\rm C}}$etaev condition for $g_\nu$ coincides with the standard definition of virtual displacements for holonomic constraints, we have $\delta^{(c)}{\dot f}_\nu=0$ (Assumption $A$). Furthermore, if the commutation relations $(C_0)$ holds, then $\delta^{(v)}{\dot f}_\nu=0$ (Assumption $B$). In standard holonomic systems, all three terms in (\ref{transprulee}) vanish.

\begin{rem}
The necessity of relations $(C_0)$) arises from the consistency between $f_\nu=0$ and $\dot{f}_\nu=0$. Defining $\delta^{(v)}f_\nu = \sum \frac{\partial f_\nu}{\partial q_j} \delta q_j$, one finds $\frac{d}{dt}(\delta^{(v)} f_\nu) = \delta^{(v)} \dot{f}_\nu$ if and only if the operations $d/dt$ and $\delta^{(c)}$ commute for any function $f_\nu$.
\end{rem}

\subsubsection{Integrable constraints via an integrating factor}

\noindent
A constraint $g_\nu$ is integrable via an integrating factor $\phi_\nu({\bf q}, t)$ if $\phi_\nu g_\nu$ is exact, i.~e.
\begin{equation}
\label{gif}
\phi_\nu ({\bf q},t)g_\nu({\bf q}, {\dot {\bf q}},t)=\dfrac{d}{dt}f_\nu ({\bf q},t).
\end{equation}
This implies $g_\nu$ is linear in ${\dot {\bf q}}$ with coefficients $a_{\nu,j}$ and $b_\nu$ satisfying
\begin{equation}
\label{relif}\phi_\nu a_{\nu,j}=\dfrac{\partial f_\nu}{\partial q_j}, \qquad \phi_\nu b_\nu=\dfrac{\partial f_\nu}{\partial t}.
\end{equation}
The necessary (and locally sufficient) closure conditions for (\ref{gif}) are
\begin{equation}
\label{chiusaif}
\begin{cases}\phi_\nu \left( \dfrac{\partial a_{\nu,i}}{\partial q_j}-\dfrac{\partial a_{\nu,j}}{\partial q_i} \right) = a_{\nu, j} \dfrac{\partial \phi_\nu}{\partial q_i} - a_{\nu, i} \dfrac{\partial \phi_\nu}{\partial q_j} \\
\phi_\nu \left( \dfrac{\partial b_\nu}{\partial q_j}-\dfrac{\partial a_{\nu,j}}{\partial t} \right) = a_{\nu, j} \dfrac{\partial \phi_\nu}{\partial t} - b_\nu \dfrac{\partial \phi_\nu}{\partial q_j}\end{cases}
\end{equation}
where $\phi_\nu=1$ recovers the standard exactness conditions. Although $\phi_\nu=0$ could theoretically introduce extraneous configurations, (\ref{chiusaif}) implies that at such points the gradients of $g_\nu$ and $\phi_\nu$ are proportional, ensuring the equivalence of $g_\nu=0$ and $\phi_\nu g_\nu=0$.

\begin{prop}
If $g_\nu$ satisfies (\ref{gif}), its Lagrangian derivative ${\cal D}_j$ obeys
\begin{equation}
\label{derlagrif}
\phi_\nu {\cal D}_j g_\nu = g_\nu \dfrac{\partial \phi_\nu}{\partial q_j} - {\dot \phi}_\nu \dfrac{\partial g_\nu}{\partial {\dot q}_j}.
\end{equation}
\end{prop}

\noindent 
{\bf Proof}: Since ${\cal D}_j{\dot f}_\nu = 0$, expanding ${\cal D}_j(\phi_\nu g_\nu)$ yields
$$
\dfrac{d}{dt}\left(\phi_\nu \dfrac{\partial g_\nu}{\partial {\dot q}_j}\right) - \phi_\nu \dfrac{\partial g_\nu}{\partial q_j} - g_\nu \dfrac{\partial \phi_\nu}{\partial q_j} = \phi_\nu {\cal D}_j g_\nu + {\dot \phi}_\nu \dfrac{\partial g_\nu}{\partial {\dot q}_j} - g_\nu \dfrac{\partial \phi_\nu}{\partial q_j} = 0. \quad \square
$$

\begin{cor}
On the constraint manifold $g_\nu=0$, we have $\phi_\nu {\cal D}_j g_\nu = -{\dot \phi}_\nu \frac{\partial g_\nu}{\partial {\dot q}_j}$.
\end{cor}

\noindent
Multiplying by $\delta q_j$ and summing over $j$, we obtain:

\begin{prop}
Let $g_\nu=0$ satisfy (\ref{gif}). If the virtual displacements $\delta q_j$ satisfy the ${\check{\rm C}}$etaev condition (Assumption $A$), then
\begin{equation}
\label{derlagrsomma0}
\sum\limits_{j=1}^n {\cal D}_j g_\nu \delta q_j = 0
\end{equation}
wherever $\phi_\nu \neq 0$.
\end{prop}

\begin{exe}
Consider $g_1 = q_2{\dot q}_1 - q_1{\dot q}_2 = 0$. Here ${\cal D}_1 g_1 = 2{\dot q}_2$ and ${\cal D}_2 g_1 = -2{\dot q}_1$. Individually these are non-zero, but the sum $2({\dot q}_2 \delta q_1 - {\dot q}_1 \delta q_2)$ vanishes under the Četaev condition $q_2 \delta q_1 - q_1 \delta q_2 = 0$ as it becomes proportional to $g_1$.
\end{exe}

\noindent
In summary, the class of constraints integrable via an integrating factor generalizes the properties of exact systems, ensuring that the transpositional rule (\ref{transprule}) reduces to the form 
$\delta^{(v)}g_\nu=	\sum\limits_{i=1}^n 
\frac{\partial g_\nu}{\partial {\dot q}_i}(\delta {\dot q}_i - \frac{d}{dt}\delta q_i)$
(whenever the ${\check{\rm C}}$etaev condition holds), mirroring the behavior of exact constraints.
While exactness (${\cal D}_j g_\nu = 0$) is sufficient for the validity of (\ref{derlagrsomma0}), the existence of an integrating factor provides a broader sufficient condition.
The reduced form of the transpositional rule is not exclusive to exact constraints, but extends to any system where an integrating factor allows for a consistent definition of virtual work.
Even though individual Lagrangian derivatives may not vanish, their weighted sum  remains zero.This leads to the interesting inverse problem: we may wonder whether the fulfillment of (\ref{derlagrsomma0}) under Assumption ($A$) implies that the constraint is necessarily of type (\ref{gif}). If true, this category would define the exact boundaries for the applicability of the reduced transpositional rule."


\subsection{Nonlinear Kinematic Constraints}

\noindent
While nonholonomic mechanics typically focuses on constraints linear or affine in generalized velocities (e.g., rolling disks), nonlinear constraints introduce significant theoretical complexities. In these cases, the compatibility of virtual variations and the principle of ideality—specifically regarding Četaev’s conditions—require careful handling, as standard variational principles often demand substantial modification.

\noindent
The first realization of such a model dates back to the Appell–Hamel machine \cite{hamel}, a benchmark system extensively discussed in literature \cite{liromp}. Despite the challenges of physical implementation, nonlinear restrictions arise naturally in theoretical contexts, such as constraints on velocity magnitude, parallelism or orthogonality between velocities, and the nonholonomic pendulum \cite{benentipendulum}.

\noindent
We focus on two key categories of nonlinear constraints: homogeneous constraints and those independent of spatial coordinates.

\subsubsection{Homogeneous Constraints}
A constraint $g_\nu({\bf q}, {\dot {\bf q}},t)$ is positive homogeneous of degree $p$ if $g_\nu({\bf q}, \lambda {\dot {\bf q}},t)=\lambda^p g_\nu({\bf q}, {\dot {\bf q}},t)$. 
According to Euler’s Theorem, this implies
\begin{equation}
\label{eulero0}
\sum_{i=1}^n {\dot q}_i \dfrac{\partial g_\nu}{\partial {\dot q}_i}=p g_\nu = 0 \quad (\text{on } g_\nu=0).
\end{equation}
The formal affinity between \eqref{eulero0} and the ${\check{\rm C}}$etaev condition (Assumption $A$) leads many authors to adopt the latter for homogeneous constraints \cite{flan}.

\begin{rem}
In terms of explicit functions (\ref{constrexpl}), if the constraints are homogeneous f any degree, the functions $\dot{q}_{m+\nu} = \alpha_\nu$ are homogeneous of degree $1$ with respect to the velocities $\dot{q}_k$ ($k=1, \dots, m$). 
Consequently, they satisfy Euler's homogeneous function theorem $$\alpha_\nu = \sum_{k=1}^m \frac{\partial \alpha_\nu}{\partial \dot{q}_k} \dot{q}_k$$
\end{rem}

\noindent 
The combination of  Combining $(A)$ with the Euler identity (\ref{eulero0}) yields the system
\begin{equation}
\label{condhom}
\begin{cases}\dfrac{\partial g_\nu}{\partial \dot{\mathbf{q}}} \cdot \dot{\mathbf{q}} = 0 \\\dfrac{\partial g_\nu}{\partial \dot{\mathbf{q}}} \cdot \delta \mathbf{q} = 0
\end{cases}\end{equation}
which implies that both the generalized velocities $\dot{\mathbf{q}}$ and the virtual displacements $\delta \mathbf{q}$ lie within the same vector space defined in (\ref{w}).

\subsubsection{Quadratic Homogeneous Constraints}
Consider homogeneous polynomial constraints of degree $2$, which represent a broad class of nonholonomic restrictions:
$$
g_\nu = \sum_{i,k=1}^n \gamma_{i,k}^{(\nu)}(\mathbf{q}) \dot{q}_i \dot{q}_k = 0.
$$
The partial derivatives with respect to the velocities are $\frac{\partial g_\nu}{\partial \dot{q}_j} = \sum_{i=1}^n \left(\gamma_{j,i}^{(\nu)} + \gamma_{i,j}^{(\nu)}\right) \dot{q}_i$. 
Consequently, the ${\check{\rm C}}$etaev condition $(A)$, corresponding to the second line of (\ref{condhom}), is expressed as
$$
\delta^{(c)}g_\nu = \sum_{j=1}^n \left[ \sum\limits_{i=1}^n \left(\gamma_{j,i}^{(\nu)} + \gamma_{i,j}^{(\nu)}\right) \dot{q}_i \right] \delta q_j = 0.
$$
The transpositional rule for these constraints becomes
\begin{equation}
\label{transphom}
\delta^{(v)}g_\nu = \sum_{j=1}^n \frac{\partial g_\nu}{\partial \dot{q}_j} \left( \delta \dot{q}_j - \frac{d}{dt}(\delta q_j) \right) - \sum\limits_{j=1}^n D_j g_\nu \delta q_j
\end{equation}
where $D_j g_\nu$ is the Lagrangian derivative of $g_\nu$. 
The physical interpretation of (\ref{transphom}) depends on the variational hypothesis adopted: if we assume $(C_0)$ (coomutation), then (\ref{transphom}) reduces to $\delta^{(v)}g_\nu =- \sum\limits_{j=1}^n D_j g_\nu \delta q_j$ and we expect $\delta^{(v)}g_\nu \neq 0$.
If $(AB)$ is assumed, then (\ref{transphom}) corresponds to (\ref{transpruleab}) and the standard commutation $(C_0)$ cannot be satisfied. 

\noindent
Generally, condition (\ref{derlagrsomma0}) is not an identity for quadratic constraints. According to Proposition 3, its validity would require $D_j g_\nu$ to be a linear combination of the coefficients $\frac{\partial g_\mu}{\partial \dot{q}_j}$:$$D_j g_\nu = \sum_{\mu=1}^\kappa \varrho^{(\nu)}_\mu \frac{\partial g_\mu}{\partial \dot{q}_j}$$for some functions $\varrho^{(\nu)}_\mu(\mathbf{q}, \dot{\mathbf{q}})$, which is typically not the case.

\noindent
Finally, combining assumptions $(A)$ and $(B)$ with statement $(C)$ from (\ref{abc0})—intended to reconcile vakonomic and nonholonomic methods \cite{pastore}—transforms (\ref{transphom}) into
\begin{equation}
\sum_{j=1}^n \left( D_j g_\nu - \sum_{i=1}^n W_{i,j} \frac{\partial g_\nu}{\partial \dot{q}i} \right) \delta q_j = 0
\end{equation}
where $\delta q_j$ satisfies condition $(A)$. While identifying the $n \times n$ coefficients $W{i,j}$ usually yields only one condition per constraint, \cite{llibre} suggests that each term in the parenthesis vanishes. This provides $\kappa \times n$ conditions for $W_{i,j}$ in terms of $\gamma_{i,j}^{(\nu)}$, their spatial derivatives, and the accelerations $\ddot{\mathbf{q}}$.

\subsubsection{Constraints depending only on velocities}
We examine the special case where the constraint functions depend only on velocities and time:
\begin{equation}
\label{noq}
g_\nu = g_\nu (\dot{\mathbf{q}}, t) = 0
\end{equation}
A physical example of this type is a constraint imposing the magnitude of the velocity of a point, such as $|\dot{\mathbf{q}}| = C(t)$, where $C(t)$ is a given non-negative function.

\noindent
The variations and the Lagrangian derivative simplify as follows:
$$
\delta^{(c)}g_\nu = \frac{\partial g_\nu}{\partial \dot{\mathbf{q}}} \cdot \delta \mathbf{q}, \qquad
\delta^{(v)}g_\nu = \frac{\partial g_\nu}{\partial \dot{\mathbf{q}}} \cdot \delta \dot{\mathbf{q}}, \qquad
D_j g_\nu = \frac{d}{dt} \left( \frac{\partial g_\nu}{\partial \dot{q}_j} \right)
$$
The general transpositional relation (\ref{transprule}) then becomes
\begin{equation}
\label{transprulenoq}
\frac{\partial g_\nu}{\partial \dot{\mathbf{q}}} \cdot \delta \dot{\mathbf{q}} - \frac{d}{dt} \left( \frac{\partial g_\nu}{\partial \dot{\mathbf{q}}} \cdot \delta \mathbf{q} \right) = \frac{\partial g_\nu}{\partial \dot{\mathbf{q}}} \cdot \left( \delta \dot{\mathbf{q}} - \frac{d}{dt}\delta \mathbf{q} \right) - \frac{d}{dt} \left( \frac{\partial g_\nu}{\partial \dot{\mathbf{q}}} \right) \cdot \delta \mathbf{q}
\end{equation}
By assuming the ${\check{\rm C}}$etaev condition $(A)$, the requirement (\ref{zero2})—equivalent to the vanishing of the Lagrangian derivative sum (\ref{derlagrsomma0})—takes the form
\begin{equation}
\label{zero2noq}\frac{d}{dt} \left( \frac{\partial g_\nu}{\partial \dot{\mathbf{q}}} \right) = \sum_{\mu=1}^\kappa \varrho_\mu \frac{\partial g_\mu}{\partial \dot{\mathbf{q}}}
\end{equation}
where the coefficients $\varrho_\mu$ are functions of $(\dot{\mathbf{q}}, t)$. If hypothesis $(B)$ also holds, relation (\ref{transprulenoq}) implies
$$
\frac{\partial g_\nu}{\partial \dot{\mathbf{q}}} \cdot \left( \delta \dot{\mathbf{q}} - \frac{d}{dt} \delta \mathbf{q} \right) = \frac{d}{dt} \left( \frac{\partial g_\nu}{\partial \dot{\mathbf{q}}} \right) \cdot \delta \mathbf{q}.
$$
It follows that the standard commutation rule $(C_0)$ is satisfied if and only if the constraints verify the condition (\ref{zero2noq}).
\begin{rem}
The condition $\frac{\partial g_\nu}{\partial \dot{\mathbf{q}}} \cdot \delta \dot{\mathbf{q}} = 0$ is a central component of Jourdain's Principle \cite{jourdain}, which applies to general constraints $g_\nu(\mathbf{q}, \dot{\mathbf{q}}, t)$. The validity of the commutation relation within the framework of this principle is further discussed in \cite{papajourdain}.
\end{rem}


\section{Conclusions and Final Remarks}
In this study, we have provided a rigorous framework for evaluating the consistency of nonholonomic mechanics by analyzing the interplay between the ${\check{\rm C}}$etaev condition $(A)$, the total variation of constraints $(B)$ and the commutation relations $(C)$. By utilizing the transpositional rule as a bridge between these formalisms, we have clarified the role of these fundamental identities in the derivation of the equations of motion.

\subsection{Summary of geometric and analytical results}

\noindent
The core findings of our analysis, rooted in the geometric characterization provided by Propositions 2.4 and 2.5, can be summarized as follows:
\begin{itemize}
	\item{} Structure of variation spaces. Proposition 2.4 establishes that while virtual displacements $\delta \mathbf{q}$ belong to the subspace $\mathbb{V}^{(n-\kappa)} = \ker B$, the consistent velocity variations $\delta \dot{\mathbf{q}}$ reside in an affine space $\mathbb{A}^{(n-\kappa)}$. This distinction is fundamental for the correct application of variational principles: the simultaneous validity of conditions ($A$) and ($B$) is mathematically restrictive and generally incompatible with standard commutation relations ($C_0$).
	\item{} Limits of commutativity. Proposition 2.5 clarifies the role of the Lagrangian derivative matrix. Commutation between the operators $\delta$ and $d/dt$ is not a general property but depends strictly on the condition $\sum_i D_i g_\nu \delta q_i = 0$. If this identity fails, the paths of variation and time-differentiation diverge into disjoint affine spaces, highlighting a fundamental divergence between variational (vakonomic) and differential (nonholonomic) methods.
	\item{} Linear vs.~nonlinear dynamics. For linear kinematic constraints, the ${\check{\rm C}}$etaev condition remains the most consistent approach. However, for nonlinear homogeneous constraints, the transpositional rule serves as a diagnostic criterion, allowing for a systematic determination of the structural requirements for consistent variations.
\end{itemize}	

\subsection{Comparative analysis and physical insight}
A key contribution of this work is the critical comparison between our minimal geometric approach and the more restrictive assumptions found in recent literature, such as the coefficients $W_{i,j}$ used in certain vakonomic models.
Our analysis demonstrates that the transpositional rule (\ref{transpruleab}) is a more economical and transparent assumption than the $\kappa \times n$ conditions proposed in \cite{llibre}. While the latter implies our results, the converse does not hold. Given that the conditions in \cite{llibre} impose significantly more constraints without a clear physical justification, we argue that our derivation offers a more robust foundation for non-classical dynamics.

\subsection{Future perspectives}
The deeper understanding of these mathematical underpinnings opens significant avenues for theoretical and applied mechanics. Future investigations will focus on:
\begin{itemize}
	\item[1.]
the physical interpretation of the $W_{i,j}$ and $\varrho_\mu^{(\nu)}$ coefficients in highly nonlinear scenarios, 
\item[2.] the application of these derived necessary conditions to advanced engineering and robotic systems, where non-classical and nonlinear constraints are increasingly prevalent, 
\item[3.] the reconciliation of the vakonomic variational method with the d'Alembert-La\-gran\-ge approach through the lens of statement ($C$).
\end{itemize}


{\small
{\em Authors' addresses}:
{\em Federico Talamucci}, University of Florence, Italy 

 e-mail: \texttt{federico.talamucci@\allowbreak unifi.it}.
}

\end{document}